\documentclass[11pt]{article}
\usepackage{aaspp4,lscape}

\def\gtsima{$\, \buildrel > \over \sim \,$}
\def\ltsima{$\, \buildrel < \over \sim \,$}
\def\simgt{\lower.5ex\hbox{\gtsima}}
\def\simlt{\lower.5ex\hbox{\ltsima}}

\begin{document}

\title{CCD Photometry of the Globlular Cluster NGC\,5986
and its Post-Asymptotic-Giant-Branch and RR Lyrae Stars }
 
\bigskip
 
\author{David R.~Alves, \ Howard E. Bond\altaffilmark{1},}

\affil{Space Telescope Science Institute, 
3700 San Martin Dr., Baltimore, MD 21218; \\
{\tt alves, bond@stsci.edu} }

\author{Christopher Onken\altaffilmark{2}}
\affil{Department of Astronomy, University of Minnesota,
116 Church Street SE, Minneapolis, \\
MN 55455; {\tt onken@sps.spa.umn.edu}  }

\altaffiltext{1}{Guest observer at Cerro Tololo Inter-American 
Observatory, National Optical Astronomy Observatories, operated
by AURA under contract with NSF.}

\altaffiltext{2}{1999 summer student at the
Space Telescope Science Institute.  Now affiliated
with the Department of Astronomy, The Ohio State University.}

\bigskip
\bigskip

\begin{abstract}

We have obtained new CCD
{\it BV\/} photometry of the 
little-studied southern Galactic globular cluster NGC\,5986,
including light curves of 5 of its RR~Lyrae variables.
The cluster's red giant branch bump is detected for the
first time at $V = 16.47 \pm 0.03$.  We derive
a reddening and true distance modulus of
$E(B-V)$~=~0.29~$\pm$~0.02 and $(m-M)_{0}$~=~15.15~$\pm$~0.10,
respectively.
The cluster's color-magnitude diagram reveals 
a mostly blue horizontal branch, like that of M13 or M2,
and quite unlike M3; yet all of these clusters have nearly
identical metallicities ([Fe/H]$_{CG97}$ = $-1.35$).
We show that the RR~Lyrae variables in NGC\,5986 are
about 0.2 mag brighter on average than those in M3, an 
important exception to the often-employed,
universal $M_V$(RR)--[Fe/H] relation.
Finally, we note that NGC\,5986 contains
two luminous stars with spectral types A-F,
which are likely to be post-asymptotic-giant-branch (PAGB)
objects.  The $V$-band luminosity 
function (LF) of such yellow PAGB stars 
is a promising standard candle.  
We suggest that the LF is sharply
peaked at $M_V$(PAGB)~=~$-3.28 \pm 0.07$.

\end{abstract}
 
\clearpage
\section{Introduction}

Population~II stars evolving off the asymptotic giant branch (AGB) 
and passing through spectral
types F and A are excellent candidates for a new extragalactic standard
candle (Bond 1997, 2001). 
These ``yellow''
post-AGB (PAGB) stars are the visually brightest members of
old populations. They 
should have a narrow luminosity function (LF), because essentially a single 
main-sequence turnoff mass is feeding the PAGB region of the HR diagram.
They are also easily recognized 
because of their enormous Balmer jumps.
A zero-point calibration for the luminosities of
Population~II yellow PAGB stars may be set 
with those in Galactic globular clusters (GCs).

The aim of this paper is to obtain the color-magnitude diagram (CMD) and
distance of the 
little-studied southern GC, NGC\,5986.  This cluster is remarkable because 
it contains two candidate A-F type PAGB stars, discovered 
some years ago during a photographic grism survey (Bond 1977).
Membership of the two PAGB stars in the cluster is confirmed by their
radial velocities and by Str\"omgren photometry showing very low
surface gravities (Bond 2001).  
This hitherto obscure cluster thus
may be destined to fill a role comparable to that of the
handful of galaxies that have produced more than one Type~Ia supernova. 
Since the cluster contains two PAGB stars, questions of cluster distance and
reddening drop out; hence their observed $V$ magnitudes will provide direct 
evidence on the width of the Population~II yellow PAGB LF.

The CMD of NGC\,5986
was first studied by Harris, Racine, \& DeRoux~(1976) 
with photographic plates.
A search for variable stars in the cluster
was made by Liller \& Lichten (1978), also with photographic plates.
We present new CCD {\it BV\/} photometry of NGC\,5986,
including new light curves of the cluster's RR~Lyrae variables.
Comparisons of our modern CCD photometry are made
with the older studies.  We provide new estimates of the
cluster's reddening and distance, and we derive the luminosities 
of the two candidate PAGB stars.

In addition to our study of the PAGB stars,
it is also interesting to compare the CMD and the
RR~Lyrae stars of NGC\,5986 to those of other Galactic GCs.
Such comparisons remain an active area of research because they
are relevant to the interrelated problems of the Oosterhoff
dichotomy (Oosterhoff 1939) and the ``second-parameter'' effect
(Sandage \& Wildey 1967), with implications for
the formation of the Galaxy (e.g., Lee, Demarque, \& Zinn 1990, 1994).
For example, 
a new, third Oosterhoff class may be emerging 
(Pritzl et al.~2000), which is not
easily understood in the context of canonical stellar
evolution theory.  Also, new exceptions to 
a universal luminosity-metallicity relation for
RR~Lyrae stars have been found in both metal-rich
(Layden et al.~1999; Pritzl et al.~2000) and metal-poor
(Lee \& Carney 1999; Clement \& Shelton 1999) clusters.
We discuss NGC\,5986 in the context
of these recent developments.

\section{Observations and Data Reduction}

Observations of NGC\,5986 and standard
star calibration fields were obtained by H.E.B.\, at the
Cerro Tololo Inter-American Observatory (CTIO) 
with the 0.9\,m telescope.  
All observations 
were made with broad-band $B$ and $V$ filters
available at CTIO,
which typically changed from run to run.  Images of
the cluster were obtained at an airmass 
close to unity.
All frames were bias-corrected and flat-fielded
in the standard manner.
The different runs and
detectors are summarized as follows.

In 1988,
observations were made with the 300$\times$500 pixel
RCA5 chip that had a scale of $0\farcs494$ 
pixel$^{-1}$ and a field of view of $2\farcm5$ by $4\farcm1$.
The cluster was observed in
$B$ and $V$ at least once on each of 9 consecutive nights from 
February 22 to March 2 (twice on 2 nights).
Typical exposures in $B$ and $V$ were 90 and 45~s, respectively.
These observations are best-suited
for our variability study.  
Unfortunately, the RCA's small field of view 
limits our search for variable stars to the central region
of the cluster.  

Subsequent observations were made on the night of 1990 June 16
with the Tek4 chip trimmed 
to 500$\times$500 pixels.  The plate scale of $0\farcs454$
pixel$^{-1}$ yielded a field of view $3\farcm7$ on a side.
These observations included photometric calibration data and
our longest exposures of the cluster: two frames of 1200~s 
each in $B$
and two frames of 300~s each in $V$.

The most recent observations 
were made on the nights of 1997 June 1
and 1998 August 26 with
the 2048$\times$2048 pixel
Tek\,2K (``No.~3'') CCD chip that had a sampling of $0\farcs396$
pixel$^{-1}$ and a field of view $13\farcm5$ on a side.  
These frames
covered a much larger sky area and thus yield the best statistics 
on the cluster giant branch and horizontal branch
in the CMD.  Exposures were
30 and 20~s in 1997 and
75 and 45~s in 1998, in the $B$ and $V$ filters, respectively. 
Photometric calibration data were obtained on both nights.

\subsection{Photometric Calibration}

As noted above, calibration frames were obtained in
1990, 1997, and 1998.  These consisted of observations of Landolt (1992)
standard-star fields repeated at different airmasses during each night.
Photometry of the standard stars was
obtained with DAOPHOT (Stetson 1987) with a
large aperture radius ($\sim 8\farcs$0).  Growth curves indicate
that the aperture magnitudes closely measure the 
total observable flux.  For each night, we 
photometered $\sim$ 20 standard stars with known colors
ranging over $(B-V) \sim -0.2$ to 1.5 and visual magnitudes
of $V \sim 10$ to 12 (1990) or $V \sim 12$ to 16 
(1997 \& 1998).

Our aperture magnitudes and the known standard system magnitudes of
Landolt (1992) were then used to derive coefficients of the transformation
equations of the form:
\begin{equation}
B \ = \ b \ + \ c_0 \ + \ c_1 (B-V) \ + \ c_2 (X - 1.0), 
\end{equation}
\begin{equation}
V \ = \ v \ + \ d_0 \ + \ d_1 (B-V) \ + \ d_2 (X - 1.0),
\end{equation}
where $b$ and $v$ are the aperture magnitudes, $B$ and $V$ are
standard magnitudes, and $X$ is airmass.  The 1997 data showed
significant scatter which we attribute to nonphotometric conditions;
they are not considered further except for differential
variable star photometry.
The 1990 and 1998 data yielded best-fit coefficients of 
($c_0$,$c_1$,$c_2$) = ($-4.546$,$+0.150$,$-0.190$) and
($-2.997$,$-0.112$,$-0.250$), and ($d_0$,$d_1$,$d_2$) =
($-3.921$,$-0.032$,$-0.140$) and ($-2.678$,$-0.004$,$-0.120$),
respectively.  The residual standard deviation of each fit
is typically 0.025 mag.  The solution residuals show no
apparent trends with standard magnitude, color, or airmass.


\subsection{Cluster Photometry}

We employed DAOPHOT/ALLSTAR (Stetson 1987) to 
identify stars and derive profile-fitted
photometry from the 1990 and 1998 frames of NGC\,5986.
Star lists were assembled in the usual manner, i.e.~by iteratively 
searching star-subtracted images
for more stars and concatenating these lists.  Model
point-spread functions (PSFs) were constructed 
with 7 and 25 isolated stars on
each of the 1990 and 1998 frames, respectively.
Mean aperture corrections for each frame
were derived with these same sets of stars.
The aperture corrections are 
uncertain at the $\sim$2\% and 1\% levels in the 1990 and 1998 data,
respectively.  This difference can be attributed to crowding.
After applying aperture corrections, 
each pair of $b$ and $v$
instrumental magnitudes was calibrated according to
Eqs.~1 \& 2.  We employ
instrumental colors, then iterate until the
solutions converge.

As a check of our calibration,
stars were identified in both the 1990 and 1998
$B$ and $V$ frames.  Comparisons of the 50 brightest 
matched stars show that
systematic zero-point offsets in each color
are less than 2\%, consistent with the
calibration uncertainties estimated above.
It is noted that the faintest stars identified in the 1990 frames
have $V \sim$ 18.5,
while the 
1998 frames yield a detection limit of $V \sim$ 20.  
This suggests that $V \sim$ 18.5 is the confusion limit 
in the crowded cluster center, which we confirm to be the
case in the 1998 data as well.  Since the photometry of faint stars
near the cluster center
is confusion limited, the 1990 
frames do not yield a significant improvement
in photometric accuracy relative to the 1998 frames
despite the longer exposures.

For the remainder of this work, the
$BV$ photometry of NGC\,5986 from the 1998 run
serves as our fiducial calibration.  Over 6000 stars were detected
in both the $B$ and $V$ frames.
These 1998 calibrated photometry data are summarized in Table~1
(the complete table is available only in the electronic
edition of {\it The Astronomical Journal\,}).  Note that the center of
the cluster is located at {\tt X}=1025, {\tt Y}=1025 in the pixel
coordinates provided, which can be used to define different
radial distance cuts for CMD construction.  
The image is available from the authors upon request.
Additional photometry
of the cluster obtained for the variable star search
is described in \S2.3.  

In Figure~1,
we compare our calibrated photometry 
to that of the Harris et al.~(1976)
photographic study for $\sim$25 stars in common.
The Harris et al.~(1976) photometry was calibrated with the
photoelectric standard-star sequence of White (1971).
Two of the White (1971) standard stars are also compared in Figure~1.
Despite the considerable scatter of the Harris et al.~(1976)
data, we find that our zero points agree to within
$\sim$0.05 mag, which we consider acceptable. 
The difference may suggest a modest degree of
flux contamination from faint neighboring stars in the 
photoelectric standard-star sequence
of White (1971).

The photographic $B$ photometry 
of cluster variable
stars by Liller \& Lichten~(1978) was calibrated relative to the 
Harris et al.~(1976) data.  In addition, their
magnitudes were estimated ``by eye'' in several instances.  Thus
we consider their photometry to be of limited use as a check of
our calibration.
Kravstov et al.~(1997) also presented
photographic photometry of NGC\,5986 which was calibrated
with the photoelectric standard star sequence of Alcaino (1984).  
Unfortunately, 
none of the Alcaino (1984) standard stars lie within our field of view.
Finally, Rosenberg et~al.~(2000) recently presented a $VI$ color-magnitude
diagram for NGC\,5986 derived from CCD data as part of a large
catalog of globular cluster CMDs. 
A visual comparison of their CMD
with ours (see \S3) suggests that no large zero-point discrepancy exists.
Judging by the cluster giant branch and the 
blue horizontal branch, our CMD shows significantly less scatter
than that obtained by
Rosenberg et al.~(2000; see their Fig.~20), although their CMD
reaches a fainter limit.  Rosenberg et al.~suggest that 
NGC\,5986 is affected by differential
reddening.  However, we do not see strong evidence for this in our
CMD, which is constructed from stars only near the cluster center.

Further discussion of the CMD of NGC\,5986 is deferred to \S3, after
the variable star content of the cluster is characterized.

\subsection{Search for Variable Stars}

As mentioned above, Liller \& Lichten (1978) searched for
variable stars in NGC\,5986 with photographic plates.  They
identified 9 periodic variables; all are RR~Lyrae variables.  
Of these,
2 are overtone pulsators (RRc; but one is actually a foreground
star) and 7 are fundamental mode
pulsators (RRab).  The cluster also contains one bright, red
semiregular variable and one severely blended variable star
whose period was not determined.

The field of view of our 1988 RCA data
includes the semiregular (V4), and the blended
variable (V10).  It also includes 5 of the RRab (V1, V2, V6, V9,
and V11), and one RRc (V12); the latter
is believed to be a foreground star and not a member
of NGC\,5986.   The known variables lying outside of our
field of view are V3, V8 (both are RRab),
and V7 (an RRc).  Note that V5 is actually a nonvariable 
star (Liller \& Lichten 1978). 

Given our modern CCD data, 
it seemed worth searching for new, possibly overlooked, variable
stars in NGC\,5986.  In addition, we decided to experiment with
a newly developed
``difference imaging'' analysis package.  Difference imaging is
well suited to the task of finding variable stars in crowded fields.
In this type of analysis, all nonvariable stars are
subtracted away in each frame, leaving isolated (and thus easy to 
photometer) images of only those stars 
which have changed brightness relative 
to a fiducial frame (e.g., Alard 2000). 

Our difference imaging analysis is briefly summarized as follows.
First, closely spaced observations on the 
same nights were averaged to increase signal to noise.
This yielded 10 $V$ frames and 9 $B$ frames over the 9 nights
in 1988.  
Source lists for each frame were generated with SExtractor 
(Bertin \& Arnouts 1996), frame-to-frame
coordinate transformations were derived with
the method of similar triangles (Groth 1986), and then all frames were
transformed to a common coordinate system\footnote{The image ``remapping''
was accomplished with code written by J. Tonry.}.
Next, $B$ and $V$ reference images
were created by taking the medians of the transformed frames.
Each frame was then subtracted from the appropriate reference image with
the ISIS program (Alard 2000).  We employed DAOPHOT 
to search the subracted frames for 3$\sigma$,
positive or negative, star-like images.  In this manner, all of the 
known variables were recovered as 3$\sigma$ detections in both $B$ and $V$,
and in at least 3 of the 9 pairs of $BV$ frames over the 9 nights.
No other stars met these criteria.
A few dozen other candidate variables (i.e., 3$\sigma$ detections in
1 or 2 pairs of $BV$ frames) were carefully inspected, 
but each could be attributed
to photon noise (i.e., subtractions of very bright stars) or 
defects in the image data.  
We estimate our detection limit for
variability to be $\sim$0.01 mag over a $\sim$10 day period.
Thus, based on this difference imaging analysis,
we conclude that the variable star survey of NGC\,5986
by Liller \& Lichten (1978) was complete in the cluster center. 

We then re-reduced the RCA frames 
with DAOPHOT/ALLSTAR in order to obtain
PSF-fitted photometry in a manner consistent with 
the 1990, 1997, and 1998 reductions (see \S2.2).
For each photometry list, stars in common with the
calibrated 1998 $BV$ data were used to derive transformations from the
instrumental ALLSTAR-reported magnitudes
directly to calibrated $B$ or $V$.
These equations included only a zero-point and color term.  The color
coefficients so obtained were consistent with those found with the
standard-star calibration data (see \S2.1).  
We estimate that each photometric 
measurement is transformed 
to the Johnson $BV$ system with an 
uncertainty of about 0.01 mag.
In total, we made
up to 12 two-color photometric measurements 
of each variable, spanning a 10 year period.
These photometric data are summarized in Table~2.

The $B$ and $V$ light curves of the 5 cluster-member RR~Lyrae stars
are shown in Figure~2, including 
the Liller \& Lichten (1978) photographic $B$ data.
No zero-point offsets have been applied
to the photographic data.
Since our new
data extend the time baseline of observations 
by an additional $\sim$15 years, we rederived the 
pulsation periods from
the combined CCD and photographic $B$-band light curves
with the supersmoother
period-finding code (Reimann 1994).
By inspection of the
period-folded light curves (see Figure~3), we 
estimate minimum and maximum
light (uncertainty of $\pm$0.05 mag), which also yields
the pulsation amplitudes ($A_{V}$ and $A_{B}$) and the
$(B-V)$ colors at minimum light.  The mean brightnesses are the
intensity-weighted averages of the calibrated CCD data (see Table~2).
These various characteristics of the light curves are summarized in Table~3.
The average magnitudes of the RRab variables in NGC\,5986
are $V$(RR) = $16.52 \pm 0.04$ and
$B$(RR) = $17.28 \pm 0.04$.

\section{The NGC\,5986 Color-Magnitude Diagram}

In Figure~3 we show the CMD for all stars lying within 3$\arcmin$ of the
center of NGC\,5986.  The time-averaged magnitudes and colors
of the cluster's variables 
(5 RRab and 1 SR) are distiguished, as are the 2 (nonvariable)
PAGB stars.   Some contamination
from field stars is clearly present in spite of the small
radius cut chosen for the CMD.
However, major features
such as the red giant branch (RGB), blue horizontal branch (HB), and 
the red-giant-branch bump (RGB-bump) near $V \approx 16.5$,
are easily identified.  
Our photometry does
not reach the cluster's subgiant branch or main sequence turn-off.  
The small number of stars associated 
with the cluster's asymptotic-giant-branch bump (AGB-bump)
are a bit confused with field star
interlopers, but this feature is located
at $V \approx 15.5$, and $(B-V) \approx 1.0$.
The RRab stars define the location of the cool, 
red portion of the cluster's HB,
which is otherwise difficult to distinguish.  The semiregular 
variable, V4, is a
bit brighter than the tip of the RGB, and is thus probably located 
very near the end (or tip) of the AGB.
The nonvariable PAGB stars (Bond 2001) 
are bluer than the RR~Lyrae variables.
Thus the blue edge of the Population~II instability strip is seen to lie
at $(B-V) \approx 0.6$ (reddened value) in NGC\,5986.  
Finally, we note that the single
yellow star which is brighter than the 2 PAGB stars in Figure~3 is not a
cluster member (Bond 2001).

The NGC\,5986 HB is predominantly blue.  
We count the blue and red HB stars by defining
appropriate regions of the CMD.
The latter are assumed to be redder than the RRab but bluer
than the RGB.  In what follows, we use $b$, $v$, and $r$ to denote the
number of blue, variable, and red HB stars, respectively.
We estimate
$b \approx$ 300 and $r \approx$ 10.  The cluster
has a total of 7 RR~Lyrae stars, but only 5 lying within 3$\arcmin$;
thus we define $v$ = 5.  Construction of 
CMDs corresponding to an equal area of sky,
but located as far away from the cluster center as our data allow,
suggest that field star contamination of the red
HB star counts ($r$) is of order a few stars, while such
contamination of the blue HB star counts
($b$) is probably negligible.  These star counts yield an HB 
morphology index (e.g., Lee, Demarque, \& Zinn 1990)
of $(b-r)/(b+v+r) \approx$ 0.92.

The clustering of stars
on the red giant branch 
at $V$(bump) = 16.470 $\pm$ 0.025 mag 
is identified as
the RGB-bump (e.g., Ferraro et al.~1999; Alves \& Sarajedini 1999).   
The location of the RGB-bump is determined by fitting the cluster's 
differential giant
branch luminosity function with a model function, as shown in Figure~4.
The fitted model function is a Gaussian 
of arbitrary height, width, and center, and a 3-parameter (quadratic) background.
The quadratic background accounts for the underlying RGB while the
Gaussian represents the RGB-bump. 
A chi-squared minimization yields
the best-fit center for the 
peak of the Gaussian and associated error, which we report above.

Guided by the cluster's RRab stars, we determine the color of 
the red giant branch at ``the level of the HB'' to be 
$(B-V)_{g} = 1.055 \pm 0.005$.  This formal error does not account
for our photometric calibration.
Our measurement of $(B-V)_{g}$ is made
by fitting a Gaussian of arbitrary height, width, and center to
a color-frequency histogram of RGB stars in the range $V$ = 16.3 to 16.9.
We adopt the best-fit center and associated error as the nominal RGB color.
Note that different $V$ magnitude cuts centered on the mean brightness
of the RRab yield consistent results.

\subsection{Metallicity}

The metallicity of NGC\,5986 is well-constrained by observations other
than our own,  
such as the integrated light of the cluster (Zinn \& West 1984), 
and spectroscopic 
measurements of the \ion{Ca}{2} triplet for 
several of the cluster's red giants (Rutledge et al.~1997).
Rutledge, Hesser \& Stetson~(1997b) 
estimate the metallicity as
[Fe/H]$_{ZW84}$ = $-1.65 \pm 0.04$ on the Zinn \& West (1984) scale, or
[Fe/H]$_{CG97}$ = $-1.35 \pm 0.04$ on the Carretta \& Gratton (1997) scale.
The color of the NGC\,5986 giant branch in the Washington
system and an assumption for the cluster's reddening
yields a consistent result (Geisler, Claria, \& Minniti 1997).
For the purposes of this work, we assume that the metallicity 
of NGC\,5986 is a known parameter, as listed above on each of the two
currently popular metallicity scales.

It is interesting to compare the known metallicity with
that derived from the periods and amplitudes of the cluster's RRab
variables.
Using the calibration of Walker \& Mack (1986) in terms of 
the $B$-band pulsation amplitude ($A_B$) and 
period, 
we find [Fe/H]$_{ZW84}$ = $-2.26 \pm 0.06$.  The calibration
of Alcock et al.~(1999) in terms of $A_V$ and period yields
[Fe/H]$_{ZW84}$ = $-2.33 \pm 0.08$.   These estimates are clearly
at odds with the metallicity derived from the spectroscopy of giants
and the integrated cluster light, which
suggests a systematic difference between the RRab 
variables in NGC\,5986 and
the calibrating RRab stars.  This will be discussed further in \S4.

\subsection{Reddening}

We first
consider two published results for the reddening toward NGC\,5986.
Zinn's (1980) analysis of the
cluster's integrated
colors yields $E(B-V) = 0.29 \pm 0.05$, while the reddening maps
of Burstein \& Heiles (1982) indicate $E(B-V) = 0.27 \pm 0.10$.
The stated uncertainties in these two measurements
were assigned by us after a careful review of the data.

We also calculate the reddening with our
new $BV$ photometry. 
First, the colors of the cluster's RRab variables at minimum light
yield an estimate of the reddening;
this is known as Sturch's (1966) method.  Employing the calibration from
Walker (1990), and the period and color data from Table~3, we
estimate $E(B-V) = 0.36 \pm 0.05$.  
The uncertainty here is dominated by
the error of the mean color at minimum light 
for the 5 RRab listed in Table~3.
Next, using $(B-V)_{g}$ as derived above,
[Fe/H]$_{CG97}$ 
from Rutledge et al.~(1997b), and the calibration of
Ferraro et al.~(1999) for the intrinsic color of the giant
branch, $(B-V)_{0,g}$, as a function of [Fe/H]$_{CG97}$
(their Eqn.~4.15), we find 
$E(B-V) = 0.26 \pm 0.04$.  
This uncertainty is dominated by the Ferraro et al.~(1999)
calibration.  

For the
remainder of this work, we adopt the weighted average
of the four reddening estimates summarized above:
$E(B-V) = 0.29 \pm 0.02$.

\subsection{Distance}

In a cluster like NGC\,5986, which has a predominantly
blue HB and only a handful of RR~Lyrae variables, it is
difficult to determine the precise level of the
zero-age horizontal branch (ZAHB; e.g., Ferraro et al.~1999).
Since the RGB-bump is not subject to horizontal-branch
evolutionary effects,
we calculate the distance to NGC\,5986 with this feature.
Adopting a standard reddening law
($R_V$ = 3.1; Cardelli, Clayton, \& Mathis 1989), 
the dereddened brightness of the RGB-bump is 
$V_0$(bump) =  15.570 $\pm$ 0.065.  This uncertainty accounts
for uncertainties in the reddening and our measurement of $V$(bump).  
The calibration
of Ferraro et al.~(1999) in terms of [Fe/H]$_{CG97}$ predicts
$M_V$(bump) = 0.425 $\pm$ 0.08, where the uncertainty represents the
standard deviation of the calibration combined with the
uncertainty in [Fe/H]$_{CG97}$ for NGC\,5986 (see \S3.1).
Thus we arrive at true distance modulus of 
$(m-M)_0 = 15.15 \pm 0.10$, or a distance of 
$10.72 \pm 0.50$ kpc.

At this distance, 
the RRab variables in NGC\,5986 have a mean absolute visual
magnitude of $M_V$(RR) = 0.47 $\pm$ 0.11.  For comparison,
the calibration\footnote{Chaboyer et al.~(1998)
find $M_V$(RR)\ = \ 
0.23($\pm$0.03)\,$\cdot$\,([Fe/H]$_{ZW84}$ + 1.9) \ + \ 0.39($\pm$0.08).}
of Chaboyer et al.~(1998)
predicts $M_V$(RR) = 0.45.   Thus the distance
to NGC\,5986 derived from the RR~Lyrae variables and this calibration
would be in good agreement with our estimate based on the RGB-bump.

\section{The RR~Lyrae Variables in NGC\,5986}

The RRab stars in the 
Bailey diagram ($\log P$ vs.~$A_V$;
see Table~3) are shown in Figure~5.  We compare the RRab stars in
NGC\,5986, the one RRab star in M15 (Pike \& Meston 1977),
and the RRab stars in M3 from 
Kaluzny et al.~(1998) and Carretta et al.~(1998).
Fiducial ``ridge'' lines for the RRab variables in 
M3 and M15 are taken from
Alcock et al.~(1999).  These ridge lines illustrate the
use of the Bailey diagram as a diagnostic of metallicity.
RRab stars with lower metallicities have longer periods
at a fixed amplitude (i.e., M15 is more metal-poor
than M3).  However, despite having nearly
identical metallicities, the RRab stars in NGC\,5986 and
M13 are unlikely to be drawn from the same distribution as
the RRab in M3 in this diagram.  
The former lie at noticeably longer periods.  This systematic
period shift explains 
the failure of the 
periods and pulsation amplitudes 
of the RRab stars in NGC\,5986 
to predict the metallicity of the cluster (see \S3.1).

Lee \& Carney (1999) have pointed out a similar problem with the
RRab variables in M2 (also in a comparison with those in M3).  Thus
the RRab stars in 3 clusters (NGC\,5986, M2, and M13) have systematically
longer periods than those in the cluster M3, yet all of these
clusters have nearly the same metallicity.  
It is interesting that the 3 clusters with the
long-period RRab variables all have blue HBs,
$(b-r)/(b+v+r)~\approx$~0.9, while M3 has
$(b-r)/(b+v+r)~\approx$~0.1.    Clement \& Shelton (1999)
have argued that the location of cluster RRab stars in the
Bailey diagram is not a unique function of metallicity.
They suggest that the
period-shift effect seen here is due to the higher luminosities
of the RRab stars, which are in a relatively more advanced 
evolutionary state. 

We compare the luminosities of the RRab variables in NGC\,5986
and M3 as follows.  From Lee \& Carney (1999), who reanalyzed the
M3 cluster photometry of Ferraro et al.~(1997) and the M3 RR~Lyrae 
photometry
of Carretta et al.~(1998), the average brightness of the
RRab stars is $V$(RR)~=~$15.665~\pm~0.001$.  
Ferraro et al.~(1997) identified the RGB-bump in M3, and found
$V$(bump)~=~$15.45~\pm~0.05$.
The reddening toward M3 is low, $E(B-V)$ = 0.01;
thus the dereddened brightnesses are 
$V_0$(RR) = $15.63$ and $V_0$(bump) = $15.42$.
Rutledge et al.~(1997b)
give [Fe/H]$_{CG97}$ = $-1.34 \pm 0.04$ for M3.
Therefore, the RGB-bump luminosity in
M3 should be the same as in NGC\,5986, or
$M_V$(bump) = 0.425.
The true M3 distance modulus is then
$(m-M)_0 = 15.00 \pm 0.10$, and the mean absolute visual magnitude
of the RRab stars is $M_V$(RR) = 0.63 $\pm$ 0.11.  
This compares
with $M_V$(RR) = 0.47 $\pm$ 0.11 found
for NGC\,5986.  Assuming that $M_V$(bump) is the same for both
M3 and NGC\,5986, the relative brightness difference between the
two groups of RRab stars 
is $\Delta V$(RR)~=~0.17~$\pm$~0.07, where $\Delta V$
is defined in the sense of M3 $-$ NGC\,5986.

As a check, the luminosity difference can also be inferred from the
period shift in the Bailey diagram.  
We estimate that the RRab variables in
NGC\,5986 are shifted by
$\Delta \log P \approx -0.07$ dex relative to the M3 ridge line
shown in Figure~4.  Following Lee \& Carney (1999), 
if the average masses
of the RR~Lyrae stars in each cluster are the same, 
the pulsation equation 
implies a luminosity difference of 
$\Delta \log (L/L_{\odot}) \approx -0.08$~dex
for this period shift.  
If the bolometric corrections are the same
in each cluster, this luminosity difference
corresponds to $\Delta V$(RR)~$\approx$~0.20~mag,
in good agreement with our estimate above.  This agreement also 
supports the assumption that the masses of RRab stars in M3 and NGC\,5986
are the same.
In summary, the RRab stars in NGC\,5986 appear to be about 0.2 mag brighter 
on average than those in M3.  

Our comparison of the RRab stars in NGC\,5986 and M3 
demonstrates an unambiguous
exception to a universal $M_V$(RR)--[Fe/H] relation.  A similar
conclusion was reached by Lee \& Carney (1999) and Clement \& Shelton (1999)
based on their analyses of the RR~Lyrae stars in other metal-poor clusters.
Exceptions to a universal $M_V$(RR)--[Fe/H] 
relation have also been found for 
RR~Lyrae stars in metal-rich GCs 
(Layden et al.~1999; Pritzl et al.~2000).  
It is not clear that the exceptions in the metal-poor and metal-rich
clusters can be explained by the same physical process.
For the case of the NGC\,5986,
we suggest that the RRab stars have evolved off of the
blue ZAHB and into the instability strip at higher luminosities, 
while those in M3 are on average less evolved, closer
to their ZAHB locations, and fainter.

\section{The PAGB Stars in NGC\,5986 }

The use of A-F spectral type (``yellow'') post-AGB stars 
as Population~II standard candles has been championed by 
Bond and collaborators 
(e.g., Bond 1997, Bond \& Fullton 1997, Bond 2001).
As discussed in \S1, the two candidate PAGB stars in
NGC\,5986 allow a test of the assertion that the
$V$-band luminosity function (LF) of Population~II yellow PAGB
stars should be sharply peaked.   

Our new $BV$ photometry (Table~1) 
and analysis of the NGC\,5986 distance yield $M_V = -3.40$ and $-3.30$
for the stars PAGB-1 and PAGB-2, respectively.  Thus, based
on only these two stars, we suggest that the
LF of Population~II yellow PAGB supergiants 
peaks near
$M_V = -3.35 \pm 0.05$.

The bolometric corrections 
of Flower (1996) yield
$\log L/L_{\odot}$~=~3.31 and 3.21 for PAGB-1 and  PAGB-2, respectively.
The luminosity-core~mass calibration 
of Vassiliadis \& Wood (1994), which is independent of
metallicity over the range of models calculated, then
predicts masses of $M$ = 0.536 and
0.528 $M_{\odot}$, respectively.  Thus the typical PAGB
remnant mass in
NGC\,5986 is about 0.53 $M_{\odot}$.
This is in fair agreement with the average mass of white dwarfs
in GCs and the halo field, $M = 0.50 \pm 0.02 M_{\odot}$, 
as recently reviewed by Alves, Bond, \& Livio (2000,
and references therein).


For comparison,
we estimate that the semiregular variable V4 in NGC\,5986
has $\log L/L_{\odot}$ = 3.44, which is higher than 
both PAGB stars.  In this case, Whitelock's (1986) period-luminosity
calibration implies
a period of about 300 days, which would be
one of the longest periods known for semiregulars in GCs.
Our data suggest a period of order $\simgt$20 days.  
However, a period as long as 300 days seems unlikely.
If this star leaves the AGB and evolves at constant luminosity
to higher temperatures, it would appear as a yellow PAGB star with
$M_V \sim -3.85$.  We caution that 
the $BV$ bolometric correction (Flower 1996) we have
employed for this cool, metal-poor giant is large and uncertain.
In this regard, near-infrared photometry would be useful.
Radial velocity data are also needed to confirm this
star's cluster membership.

Based on only the two yellow PAGB stars in NGC\,5986,
the intrinsic scatter, or width, 
of the peak of $V$-band LF
appears to be quite narrow.  However, we are obviously dealing with
small-number statistics.  There is one other well-known
yellow PAGB star in a GC; this is ROA~24 in $\omega$~Cen, a  
confirmed cluster member.
Adopting the $BV$ photometry assembled by
Gonz\'{a}lez \& Wallerstein (1992), and
the reddening and distance modulus from
Alcaino \& Liller (1987; see also Harris 1996),
we estimate\footnote{We note that $M_V$ for ROA~24 
in Gonz\'{a}lez \& Wallerstein (1992) was incorrectly calculated
from $V$ and $(m-M)_0$ instead of $(m-M)_V$.} 
$M_V = -3.15 \pm 0.12$, $\log L/L_{\odot}$ = 3.17, and $M = 0.526 M_{\odot}$.
ROA~24 combined with the two PAGB stars in NGC\,5986 suggests that the
intrinsic width of the $V$-band yellow PAGB LF is $\sigma_V \sim$ 0.10--0.15 mag,
with a peak at $M_V(PAGB) = -3.28 \pm 0.07$.

Recently, Neely, Sarajedini, \& Martins (2000) presented a new CMD of
the GC NGC\,6144, and suggested that this cluster also contains
a yellow PAGB star.  Using
the $M_V$(RR) calibration of Chaboyer (1998) as in \S3.3, and the reddening, 
metallicity, and $V$(HB) estimates from Neely et al.~(2000), this 
candidate yellow PAGB star has $M_V = -4.04$, 
$\log L/L_{\odot}$ = 3.51, and $M = 0.56 M_{\odot}$.  
Based on this high luminosity and mass, we suggest that this star
may not be a member of the cluster.  Measurement of the Balmer jump
would confirm the PAGB nature of this star.
Radial velocity data are also needed to confirm its
cluster membership.  

The existence of {\it two} yellow PAGB
stars in NGC\,5986 argues that their lifetimes are relatively long.  
Extant stellar evolution theory cannot predict yellow
PAGB lifetimes in an absolute sense,
because the transition 
from the tip of the AGB to yellow PAGB star 
is critically dependent on ad hoc mass-loss parameterizations
(Trams et al.~1989; Vassiliadis \& Wood 1994).
This is not the case for hotter
PAGB stars, such as the nuclei of PNe, whose evolutionary
timescales are less sensitive to the mass-loss rates.
If yellow PAGB stars live 2-3$\times 10^4$ years, and assuming
each star in a GC becomes a yellow PAGB star, the total luminosity
of the GC system implies a total of $\sim$16 yellow PAGBs in GCs.
Based on the 3 known PAGB stars in the GC system (2 in NGC\,5986
and 1 in $\omega$~Cen), a minimum lifetime is $\simgt$4000 years.
However, we remind that
no systematic search of the GC system for yellow PAGBs
has yet been conducted.

The calibration of yellow PAGB stars as standard candles
will require a larger sample of objects,
probably of order a dozen stars in Galactic GCs, 
in order to accurately define the
peak and width of the $V$-band LF.  
Close-binary merger scenarios
may introduce outliers on the bright side
of what may otherwise be 
a sharply peaked LF (e.g., Alves et al.~2000). 
Nonvariable
stars lying blueward of the Population~II instability strip,
but in an evolutionary state related to the Population~II Cepheids,
may introduce outliers on both the faint and bright sides of the LF.
Indeed, some Population~II Cepheids in GCs
reach luminosities of $M_V \sim -3.5$
(e.g., Gonz\'{a}lez \& Wallerstein 1992).  
If yellow ``cousins'' of the Population~II Cepheids
significantly contaminate the PAGB LF,
we predict a correlation between the occurrence of yellow PAGBs
in GCs and the
clusters' HB morphologies, as the Population~II Cepheids 
show (Wallerstein 1970).  We have undertaken a large-scale
survey of the GC
system for yellow PAGB stars, which will help to
define observationally the true peak and width of their $V$-band LF,
and perhaps shed new light on the unusual presence of {\it two\,}
PAGB stars in NGC\,5986.

\section{Conclusion}

We have presented a new $BV$ color-magnitude diagram of
NGC\,5986, as well as new $BV$ light curves for 5 of the 
cluster's RR~Lyrae variables.  We have detected the RGB-bump
for the first time on this cluster's giant branch.  Our analysis
of the color-magnitude diagram yields new estimates for the reddening 
and true distance modulus: $E(B-V) = 0.29 \pm 0.02$ and 
$(m-M)_0 = 15.15 \pm 0.10$, respectively.  In comparison with
the globular cluster M3, which has the same metallicity as NGC\,5986,
we find that the RR~Lyrae stars in the latter are about 0.2 mag brighter.
This is an important exception to a universal $M_V$(RR)--[Fe/H]
relation in a metal-poor cluster.  Our $BV$ photometry of the
two candidate yellow PAGB stars in NGC\,5986 suggests that the
$V$-band LF of these stars is sharply peaked at
$M_V$(PAGB) = $-3.28 \pm 0.07$.

\begin{acknowledgements}
This work was partially supported by NASA grant NAG5-6821
under the ``UV, Visible, and Gravitational
Astrophysics Research and Analysis'' program.
H.E.B.~thanks the staff at Cerro Tololo for their
support over the years.
\end{acknowledgements}


\clearpage
\begin{figure}
\plotone{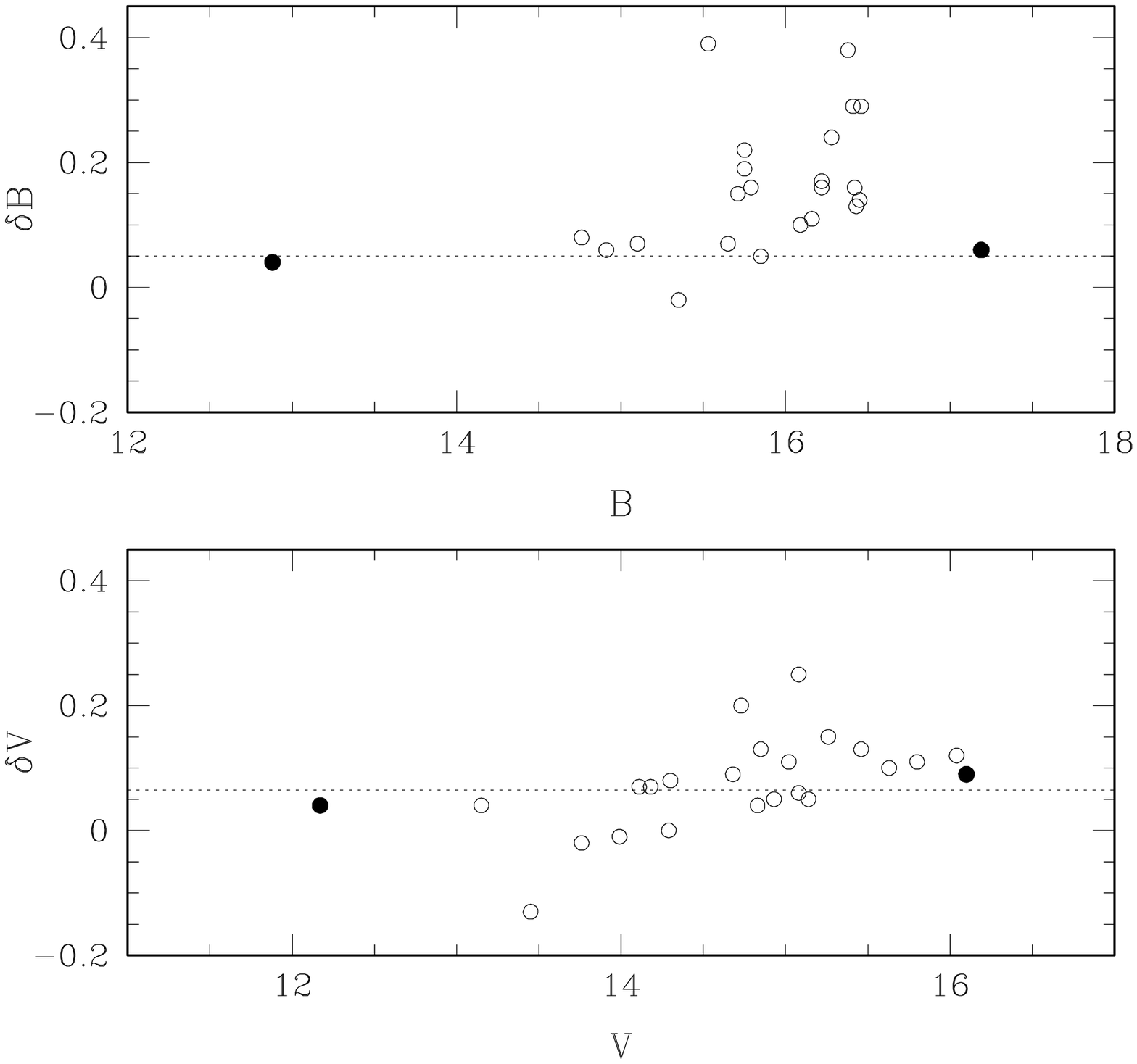}
\caption{Comparison of our calibrated photometry 
with the photographic photometry of Harris et al.~(1976;
shown as open circles) and the photoelectric standard sequence
photometry of White (1971; shown as filled circles).
In each panel, we plot the magnitude difference ($\delta B$ and
$\delta V$ = this work $-$ past work) as a function of our
calibrated magnitudes ($B$ and $V$).  The dotted lines show
the mean offsets calculated from the two photoelectric
standard stars.}
\end{figure}

\clearpage
\begin{figure}
\plotone{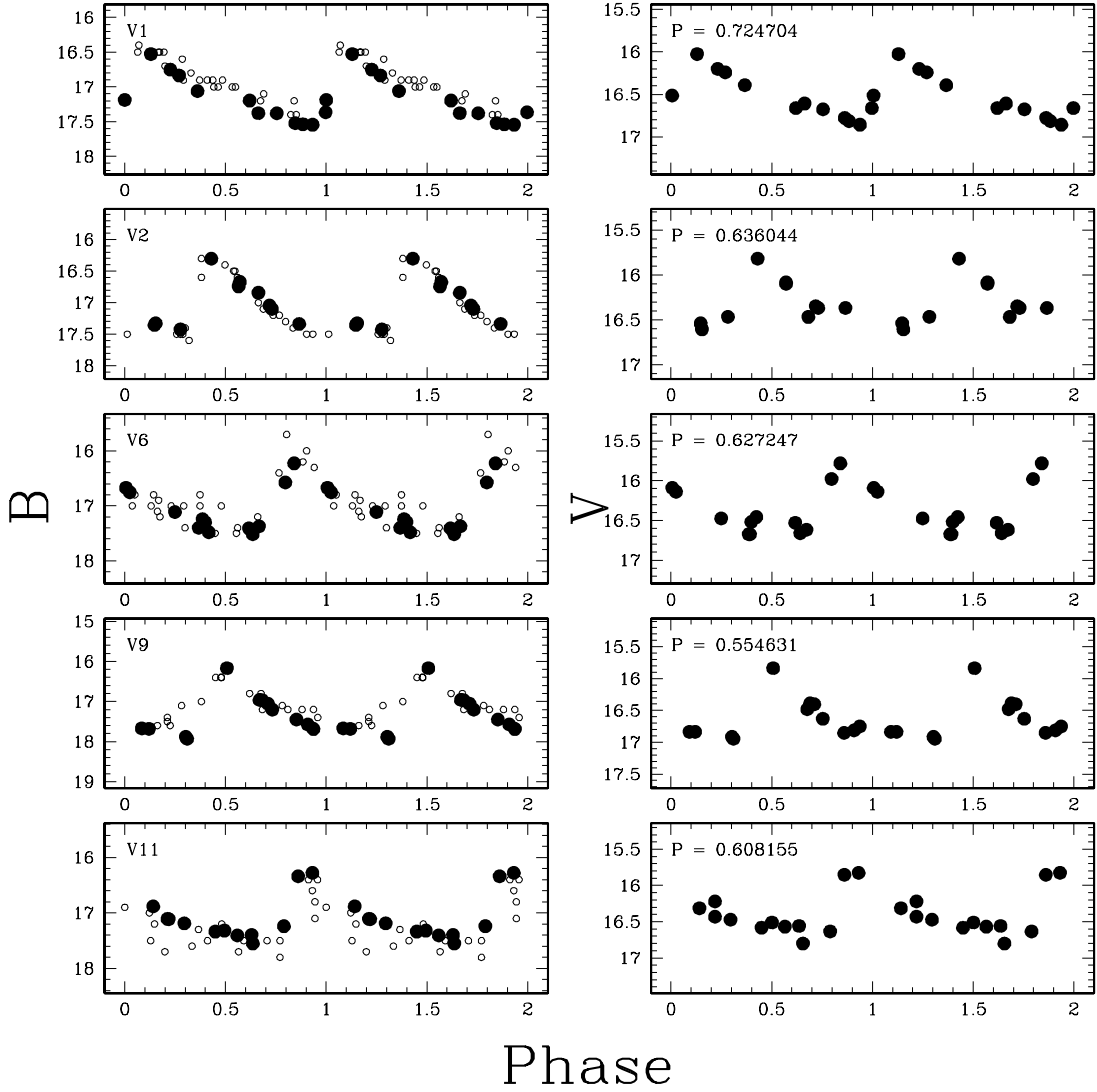}
\caption{Period-folded light curves of 5 type ab RR~Lyrae in
NGC\,5986.  The $B$ and $V$ light curves are shown in the
left and right panels, respectively.  The variable identifier
is labeled in the $B$ panel.  The period is labeled in the
$V$ panel. Our new photometric data are indicated with filled
circles.  The photographic $B$ photometric data from Liller \& Lichten
(1978) are shown with open circles.   The periods have been
refined using the combined datasets, but are in good agreement with 
the original estimates by Liller \& Lichten (1978).}
\end{figure}

\clearpage
\begin{figure}
\plotone{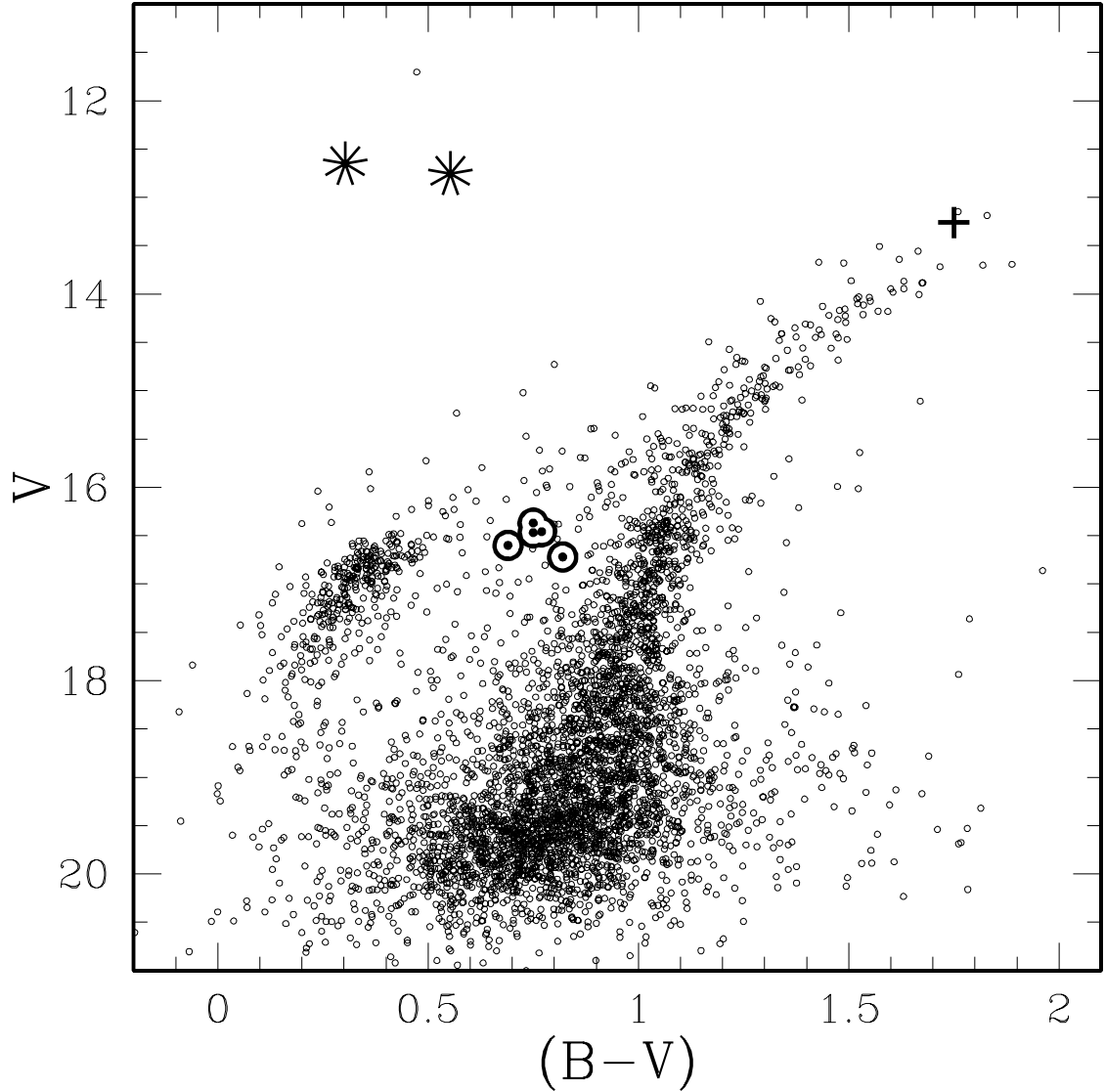}
\caption{CMD of NGC\,5986; all stars within 3$\arcmin$
of the cluster center are plotted. Five RRab stars
are shown with bulls-eyes
(intensity-weighted average positions).  The cluster's
semiregular variable is marked with a plus. 
The two PAGB stars are shown with asterisks; 
note also the one foreground star of
similar brightness and color (Bond 2000).}
\end{figure}

\clearpage
\begin{figure}
\plotone{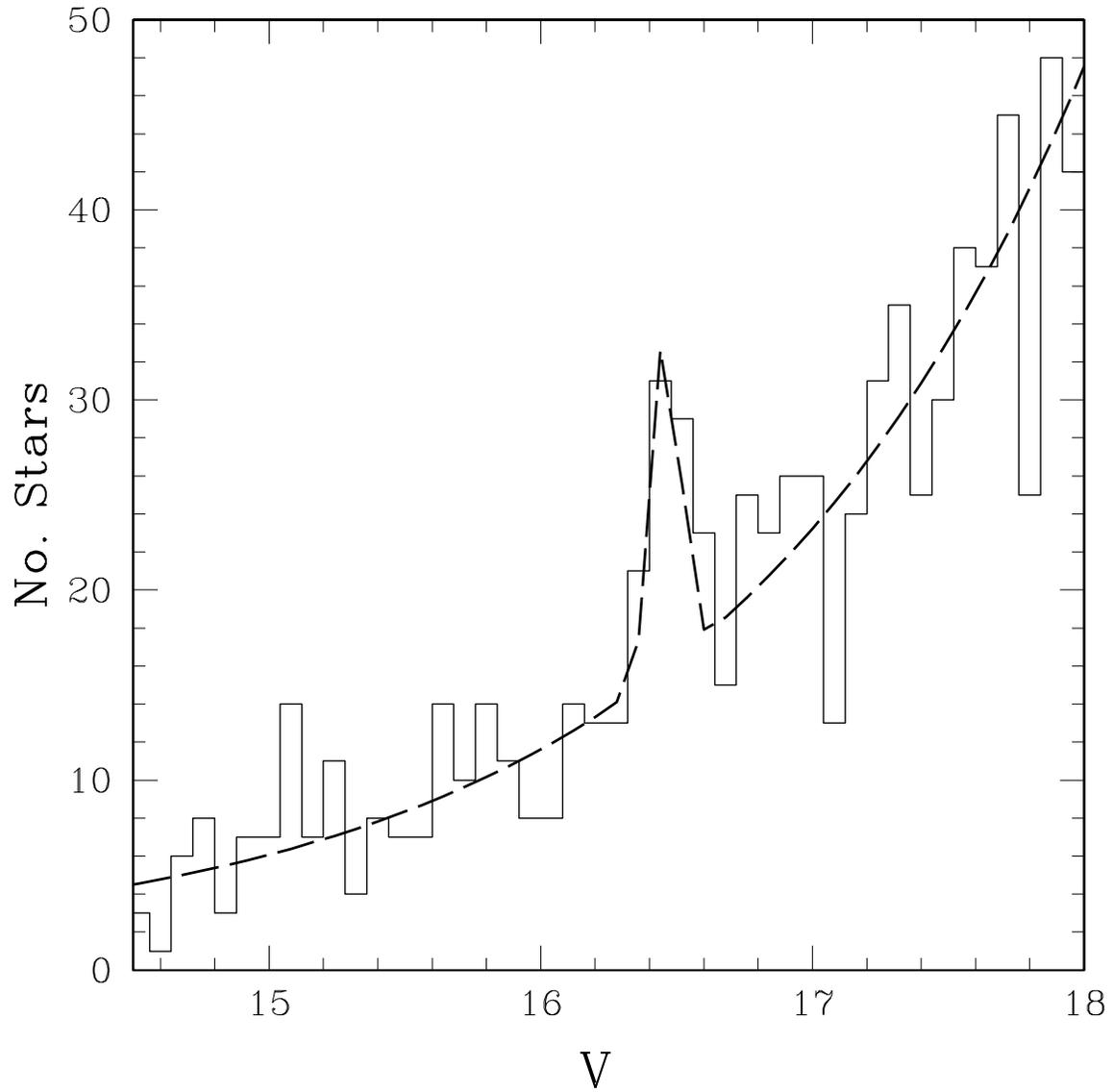}
\caption{The cluster's giant branch differential luminosity function (histogram),
with our best fit model function overplotted (dashed line).   The 
RGB-bump is peaks at $V = 16.47 \pm 0.03$. }
\end{figure}

\clearpage
\begin{figure}
\plotone{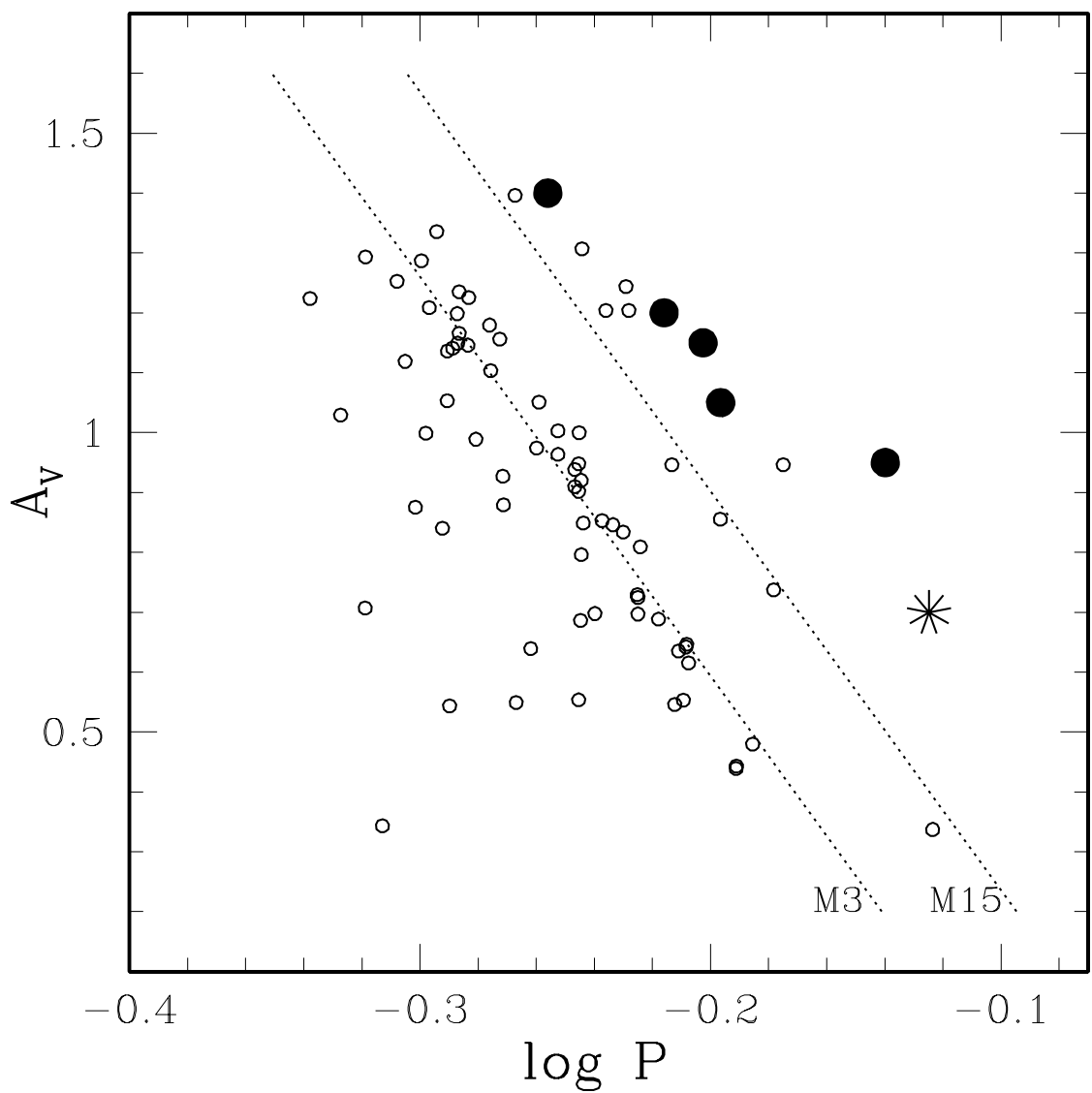}
\caption{The Bailey period-amplitude diagram.
The RRab in NGC\,5986 are plotted as bold filled circles.
We plot the RRab in M3 from Kaluzny et al.~(1998) and Carretta et al.~(1998) 
as small open circles.  The one RRab in the cluster M13
is shown with an asterisk (Pike \& Meston 1977).
Fiducial ``ridge'' lines for the RRab in the clusters
M3 and M15 are indicated with dotted lines and labeled.}
\end{figure}

\clearpage

\begin{deluxetable}{lrrccccl}
\tablewidth14cm
\tablecaption{Calibrated $BV$ Photometry of NGC\,5986 and Surrounding Field}
\tablehead{
\colhead{ID} &
\colhead{X$_{\tt PIX}$\tablenotemark{a}} &
\colhead{Y$_{\tt PIX}$\tablenotemark{a}} &
\colhead{$B$} &
\colhead{$\sigma_B$} &
\colhead{$V$} &
\colhead{$\sigma_V$} &
\colhead{Comment\tablenotemark{b}} 
}
\startdata
 1 &  874.035 &  977.380 & 12.176 & 0.008 & 11.703 & 0.014 & Star F \nl
 2 &  317.649 & 1486.862 & 12.876 & 0.019 & 12.169 & 0.015 &  \nl
 3 & 1636.164 & 1269.349 & 13.957 & 0.018 & 12.503 & 0.013 &  \nl
 4 &  242.104 & 1859.430 & 13.833 & 0.028 & 12.560 & 0.024 &  \nl
 5 & 1605.885 &  836.089 & 13.502 & 0.012 & 12.617 & 0.011 &  \nl
 6 & 1013.955 &  964.771 & 12.955 & 0.010 & 12.652 & 0.010 & PAGB-1 \nl
 7 &  966.942 & 1010.151 & 13.308 & 0.008 & 12.755 & 0.014 & PAGB-2 \nl
 8 & 1801.448 & 1977.540 & 13.682 & 0.035 & 12.879 & 0.031 &  \nl
\enddata
\tablenotetext{}{NOTE -- Table 1 is presented in its entirety in the
electronic edition of the Astronomical Journal.  A portion is shown
here for guidance regarding its form and content.}
\tablenotetext{a}{Pixel coordinates from 1998 data; the plate scale
is $0\farcs396$ pixel$^{-1}$  The center of the
cluster is at approximately X = 1025, Y = 1025.}
\tablenotetext{b}{Star F is a foreground F dwarf.  PAGB-1 and PAGB-2
are the the two PAGB candidates, denoted ``BW'' and ``BE'' in
Bond (2000).  Known variable star designations (i.e., V4) are also listed
in the complete version of this table.}
\end{deluxetable}

\clearpage

\begin{deluxetable}{lllllll}
\tablewidth11cm
\tablecaption{Time-Series $BV$ Photometry of Variable Stars}
\tablehead{
\colhead{ID} &
\colhead{$V$} &
\colhead{$\sigma_V$} &
\colhead{H.J.D.$_V$} &
\colhead{$B$} &
\colhead{$\sigma_B$} &
\colhead{H.J.D.$_B$} \\
\colhead{} &
\colhead{} &
\colhead{} &
\colhead{2,400,000+} &
\colhead{} &
\colhead{} &
\colhead{2,400,000+} 
}
\startdata
V1 & 16.200 & 0.010 & 47213.902 & 16.752 & 0.013 & 47213.898 \nl
   & 16.662 & 0.026 & 47214.910 & 17.197 & 0.039 & 47214.910 \nl
   & 16.662 & 0.022 & 47215.906 & 17.365 & 0.021 & 47215.906 \nl
   & 16.392 & 0.009 & 47216.898 & 17.060 & 0.022 & 47216.895 \nl
   & 16.676 & 0.017 & 47217.906 & 17.380 & 0.029 & 47217.906 \nl
   & 16.025 & 0.010 & 47218.902 & 16.526 & 0.012 & 47218.902 \nl
   & 16.816 & 0.020 & 47220.898 & 17.538 & 0.025 & 47220.898 \nl
   & 16.240 & 0.009 & 47221.902 & 16.836 & 0.011 & 47221.902 \nl
   & 16.608 & 0.013 & 47222.910 & 17.377 & 0.024 & 47222.910 \nl
   & 16.777 & 0.014 & 48058.641 & 17.521 & 0.015 & 48058.629 \nl
   & 16.857 & 0.030 & 50601.676 & 17.544 & 0.030 & 50601.672 \nl
   & 16.513 & 0.013 & 51052.496 & 17.187 & 0.016 & 51052.492 \nl
 & & & & & & \nl 
V2 & 16.084 & 0.019 & 47213.902 & 16.743 & 0.024 & 47213.898 \nl
   & 16.606 & 0.039 & 47214.910 & 17.327 & 0.041 & 47214.910 \nl
   & 16.348 & 0.031 & 47215.906 & 17.045 & 0.027 & 47215.906 \nl
   & 16.465 & 0.020 & 47216.898 & 17.424 & 0.056 & 47216.895 \nl
   & 16.367 & 0.033 & 47217.906 & 17.337 & 0.034 & 47217.906 \nl
   & 15.819 & 0.020 & 47218.902 & 16.303 & 0.019 & 47218.902 \nl
   & 16.099 & 0.032 & 47220.898 & 16.672 & 0.026 & 47220.898 \nl
   & 16.538 & 0.030 & 47221.902 & 17.354 & 0.037 & 47221.902 \nl
   & 16.368 & 0.027 & 47222.910 & 17.099 & 0.027 & 47222.910 \nl
   & 16.467 & 0.066 & 48058.641 & 16.843 & 0.027 & 48058.629 \nl
   & 16.334 & 0.019 & 51052.496 & 16.929 & 0.014 & 51052.492 \nl
 & & & & & & \nl 
V4 & 13.373 & 0.004 & 47213.902 & 15.122 & 0.003 & 47213.898 \nl
   & 13.310 & 0.006 & 47214.910 & 15.067 & 0.010 & 47214.910 \nl
   & 13.292 & 0.005 & 47215.906 & 15.049 & 0.007 & 47215.906 \nl
   & 13.278 & 0.006 & 47216.898 & 15.016 & 0.008 & 47216.895 \nl
   & 13.208 & 0.005 & 47217.906 & 14.984 & 0.006 & 47217.906 \nl
   & 13.207 & 0.005 & 47218.902 & 14.940 & 0.005 & 47218.902 \nl
   & 13.168 & 0.009 & 47220.898 & 14.869 & 0.011 & 47220.898 \nl
   & 13.154 & 0.007 & 47221.902 & 14.866 & 0.006 & 47221.902 \nl
   & 13.149 & 0.004 & 47222.910 & 14.840 & 0.004 & 47222.910 \nl
   & 13.368 & 0.015 & 48058.641 & 15.128 & 0.004 & 48058.629 \nl
   & 13.219 & 0.044 & 50601.676 & 14.904 & 0.019 & 50601.672 \nl
   & 13.312 & 0.012 & 51052.496 & 15.183 & 0.009 & 51052.492 \nl
 & & & & & & \nl
V6 & 16.660 & 0.024 & 47213.902 & 17.515 & 0.023 & 47213.898 \nl
   & 16.474 & 0.030 & 47214.910 & 17.112 & 0.045 & 47214.910 \nl
   & 15.782 & 0.015 & 47215.906 & 16.228 & 0.014 & 47215.906 \nl
   & 16.458 & 0.035 & 47216.898 & 17.481 & 0.040 & 47216.895 \nl
   & 16.138 & 0.015 & 47217.906 & 16.754 & 0.024 & 47217.906 \nl
   & 16.529 & 0.039 & 47218.902 & 17.412 & 0.021 & 47218.902 \nl
   & 15.978 & 0.021 & 47220.898 & 16.574 & 0.014 & 47220.898 \nl
   & 16.517 & 0.031 & 47221.902 & 17.298 & 0.018 & 47221.902 \nl
   & 16.090 & 0.015 & 47222.910 & 16.671 & 0.013 & 47222.910 \nl
   & 16.672 & 0.028 & 48058.641 & 17.399 & 0.014 & 48058.629 \nl
   & 16.617 & 0.037 & 50601.676 & 17.372 & 0.038 & 50601.672 \nl
   & 16.672 & 0.025 & 51052.496 & 17.243 & 0.023 & 51052.492 \nl
 & & & & & & \nl
V9 & 16.390 & 0.019 & 47213.902 & 16.972 & 0.018 & 47213.898 \nl
   & 15.835 & 0.025 & 47214.910 & 16.166 & 0.026 & 47214.910 \nl
   & 16.915 & 0.035 & 47215.906 & 17.881 & 0.043 & 47215.906 \nl
   & 16.837 & 0.029 & 47216.898 & 17.668 & 0.037 & 47216.895 \nl
   & 16.813 & 0.026 & 47217.906 & 17.570 & 0.055 & 47217.906 \nl
   & 16.403 & 0.020 & 47218.902 & 17.058 & 0.020 & 47218.902 \nl
   & 16.946 & 0.034 & 47220.898 & 17.931 & 0.043 & 47220.898 \nl
   & 16.836 & 0.028 & 47221.902 & 17.680 & 0.024 & 47221.902 \nl
   & 16.752 & 0.030 & 47222.910 & 17.688 & 0.039 & 47222.910 \nl
   & 16.632 & 0.016 & 48058.641 & 17.202 & 0.013 & 48058.629 \nl
   & 16.852 & 0.050 & 50601.676 & 17.452 & 0.043 & 50601.672 \nl
   & 16.480 & 0.027 & 51052.496 & 16.959 & 0.019 & 51052.492 \nl
 & & & & & & \nl 
V10 & 15.330 & 0.031 & 47213.902 & 16.537 & 0.013 & 47213.898 \nl
    & 15.276 & 0.044 & 47214.910 & 16.664 & 0.079 & 47214.910 \nl
    & 15.289 & 0.050 & 47215.906 & 16.696 & 0.065 & 47215.906 \nl
    & 15.306 & 0.053 & 47216.898 & 16.627 & 0.077 & 47216.895 \nl
    & 15.301 & 0.030 & 47217.906 & 16.598 & 0.049 & 47217.906 \nl
    & 15.300 & 0.047 & 47218.902 & 16.604 & 0.048 & 47218.902 \nl
    & 15.248 & 0.054 & 47220.898 & 16.701 & 0.066 & 47220.898 \nl
    & 15.271 & 0.063 & 47221.902 & 16.608 & 0.031 & 47221.902 \nl 
    & 15.275 & 0.036 & 47222.910 & 16.825 & 0.066 & 47222.910 \nl
 & & & & & & \nl 
V11 & 16.558 & 0.020 & 47213.902 & 17.396 & 0.032 & 47213.898 \nl
    & 16.472 & 0.035 & 47214.910 & 17.189 & 0.045 & 47214.910 \nl
    & 15.825 & 0.016 & 47215.906 & 16.277 & 0.016 & 47215.906 \nl
    & 16.568 & 0.054 & 47216.898 & 17.407 & 0.040 & 47216.895 \nl
    & 16.432 & 0.051 & 47217.906 & 17.112 & 0.030 & 47217.906 \nl
    & 15.852 & 0.027 & 47218.902 & 16.339 & 0.018 & 47218.902 \nl
    & 16.313 & 0.049 & 47220.898 & 16.883 & 0.034 & 47220.898 \nl
    & 16.633 & 0.025 & 47221.902 & 17.240 & 0.027 & 47221.902 \nl
    & 16.583 & 0.036 & 47222.910 & 17.338 & 0.031 & 47222.910 \nl
    & 16.801 & 0.035 & 48058.641 & 17.552 & 0.028 & 48058.629 \nl
    & 16.221 & 0.056 & 50601.676 & 17.108 & 0.043 & 50601.672 \nl
    & 16.510 & 0.033 & 51052.496 & 17.319 & 0.041 & 51052.492 \nl
 & & & & & & \nl 
V12 & 14.519 & 0.005 & 47213.902 & 14.929 & 0.004 & 47213.898 \nl
    & 14.449 & 0.007 & 47214.910 & 14.850 & 0.014 & 47214.910 \nl
    & 14.502 & 0.007 & 47215.906 & 14.905 & 0.008 & 47215.906 \nl
    & 14.853 & 0.011 & 47216.898 & 15.356 & 0.009 & 47216.895 \nl
    & 14.892 & 0.011 & 47217.906 & 15.486 & 0.008 & 47217.906 \nl
    & 14.868 & 0.006 & 47218.902 & 15.429 & 0.007 & 47218.902 \nl
    & 14.518 & 0.011 & 47220.898 & 14.968 & 0.012 & 47220.898 \nl
    & 14.435 & 0.009 & 47221.902 & 14.863 & 0.007 & 47221.902 \nl
    & 14.480 & 0.007 & 47222.910 & 14.884 & 0.005 & 47222.910 \nl
    & 14.700 & 0.020 & 50601.676 & 15.152 & 0.023 & 50601.672 \nl
    & 14.639 & 0.011 & 51052.496 & 15.109 & 0.010 & 51052.492 \nl
\enddata
\end{deluxetable}

\clearpage


\begin{deluxetable}{llllcccl}
\tablewidth15cm
\tablecaption{Summary of Light Curve Characteristics}
\tablehead{
\colhead{ID} &
\colhead{P (days)} &
\colhead{$A_B$} &
\colhead{$A_V$} &
\colhead{$<B>$} &
\colhead{$<V>$} &
\colhead{$(B-V)_{\rm min}$} &
\colhead{Type} 
}
\startdata
V1  & 0.724704   & 1.20        & 0.95 & 17.29 & 16.60 & 0.75 & RRab	\nl
V2  & 0.636044   & 1.60        & 1.05 & 17.12 & 16.37 & 1.00 & RRab	\nl
V4  & $\simgt$20 & 0.3         & 0.2  & 13.26 & 15.01 &      & Semiregular \nl
V6  & 0.627247   & 1.60        & 1.15 & 17.23 & 16.46 & 0.80 & RRab	\nl
V9  & 0.554631   & 2.00        & 1.40 & 17.54 & 16.72 & 0.90 & RRab	\nl
V10 & \nodata    & 0.2         & 0.1  & 16.66 & 15.29 &      & 		\nl
V11 & 0.608155   & 1.50        & 1.20 & 17.22 & 16.47 & 0.80 & RRab	\nl
V12 & 0.348834   & 0.70        & 0.50 & 15.15 & 14.66 &      & foreground RRc \nl
\enddata
\end{deluxetable}


\end{document}